\documentclass[fleqn,usenatbib]{mnras}

\usepackage{newtxtext,newtxmath}

\usepackage[T1]{fontenc}

\usepackage{graphicx}   
\usepackage{amsmath}    
\usepackage{amssymb}    
\usepackage{hyperref}

\title[X-ray AGN in post-mergers]{The X-ray View of Merger-Induced AGN Activity at Low Redshift}

\author[Secrest et al.]{
Nathan J. Secrest$^{1}$\thanks{E-mail: nathan.secrest@navy.mil},
Sara L. Ellison$^{2}$,
Shobita Satyapal$^{3}$,
\& Laura Blecha$^{4}$
\\
$^{1}$U.S. Naval Observatory, 3450 Massachusetts Ave NW, Washington, DC 20392-5420, USA\\
$^{2}$Department of Physics and Astronomy, University of Victoria, Victoria, BC V8P 1A1, Canada\\
$^{3}$Department of Physics and Astronomy, George Mason University, MS3F3, 4400 University Drive, Fairfax, VA 22030, USA\\
$^{4}$Department of Physics, University of Florida, P.O. Box 118440, Gainesville, FL 32611-8440, USA
}

\date{Accepted XXX. Received YYY; in original form ZZZ}

\pubyear{2019}

\begin{document}
\label{firstpage}
\pagerange{\pageref{firstpage}--\pageref{lastpage}}
\maketitle

\begin{abstract}
Galaxy mergers are predicted to trigger accretion onto the central supermassive black holes, with the highest rates occurring during final coalescence. Previously, we have shown elevated rates of both optical and mid-IR selected active galactic nuclei (AGN) in post-mergers, but to date the prevalence of X-ray AGN has not been examined in the same systematic way. We present \textit{XMM-Newton} data of 43 post-merger galaxies selected from the \textit{Sloan Digital Sky Survey} along with 430 non-interacting control galaxies matched in stellar mass, redshift, and environment in order to test for an excess of hard X-ray (2--10 keV) emission in post-mergers attributable to triggered AGN. We find 2 X-ray detections in the post-mergers ($4.7^{+9.3}_{-3.8}\%$) and 9 in the controls ($2.1^{+1.5}_{-1.0}\%$), an excess of $2.22^{+4.44}_{-2.22}$, where the confidence intervals are 90\%. While we therefore do not find statistically significant evidence for an X-ray AGN excess in post-mergers ($p=0.26$), we find a factor of $\sim17$ excess of mid-IR AGN in our sample, consistent with past work and inconsistent with the observed X-ray excess ($p=2.7\times10^{-4}$). Dominant, luminous AGN are therefore more frequent in post-mergers, and the lack of a comparable excess of 2--10~keV X-ray AGN suggests that AGN in post-mergers are more likely to be heavily obscured. Our results are consistent with the post-merger stage being characterised by enhanced AGN fueling, heavy AGN obscuration, and more intrinsically luminous AGN, in line with theoretical predictions.
\end{abstract}




\section{Introduction} \label{sec: Introduction}
The supermassive black holes (SMBHs) that power active galactic nuclei (AGN) appear to have been in place from the earliest epochs \citep[e.g.][and references therein]{2018Natur.553..473B}, and the ubiquitous presence of AGN throughout cosmic history makes the AGN power source, SMBH growth mechanism, and the effect of AGN on their host galaxies a central topic of study for extragalactic astrophysics \citep[for a review, see][]{2013ARA&A..51..511K}. While mergers are a central event in the build-up of galaxies over cosmic history, and are a natural mechanism for nuclear fueling, the relationship between galaxy mergers and periods of AGN activity has remained murky. On the one hand, numerous morphological studies of moderate redshift ($z\sim0.3-2.5$) AGN host galaxies have found no significant preference for mergers over isolated galaxies, suggesting that mergers do not play a significant role in the fueling of SMBHs \citep[e.g.][]{2011ApJ...726...57C, 2012ApJ...744..148K, 2012MNRAS.425L..61S, 2014MNRAS.439.3342V, 2017MNRAS.466..812V, 2019MNRAS.483.2441V}. However, an important subtlety appears to be how the AGN are selected, with obscured AGN being found more frequently in mergers \citep{2015A&A...573A.137L, 2015ApJ...814..104K, 2017MNRAS.468.1273R, 2018ApJ...853...63D}, along with radio-loud AGN \citep{2015ApJ...806..147C} and the most bolometrically luminous and/or reddened AGN \citep{2010ApJ...716L.125K, 2012ApJ...746L..22K, 2012ApJ...758L..39T, 2015ApJ...806..218G, 2016ApJ...822L..32F, 2018Natur.563..214K}.

Approaching the question in the opposite direction, studies of galaxy mergers at low redshift find a statistically significant enhanced AGN fraction \citep{2011MNRAS.418.2043E, 2013MNRAS.435.3627E, 2014MNRAS.441.1297S, 2014AJ....148..137L, 2018PASJ...70S..37G,2019MNRAS.487.2491E}, and \citet{2015MNRAS.451L..35E} have previously shown that for SDSS galaxies at $z\sim0$ the merger-AGN connection depends on selection technique (e.g. optical emission lines, mid-IR colour, radio properties). Such statistical studies of the merger-AGN connection at low $z$ were made possible by the availability of large, multi-wavelength sky surveys such as SDSS, FIRST/NVSS \citep{1995ApJ...450..559B,1998AJ....115.1693C}, and the \textit{Wide-field Infrared Survey Explorer} \citep{2010AJ....140.1868W}. To date, however, there have been only two all-sky X-ray surveys: \textit{ROSAT} \citep{1999A&A...349..389V}, which operated at very soft X-ray energies (0.1--2.5~keV), and the Neil Gehrels \textit{Swift} Observatory Burst Alert Telescope \citep[BAT;][]{2005SSRv..120..143B}, which continuously surveys the sky at very hard X-ray energies (14--195~keV). While both surveys are quite shallow and have poor angular resolution, the sensitivity of BAT to very hard X-rays makes it uniquely capable of detecting both obscured and unobscured AGN with high reliability and completeness in the local universe. This has enabled important studies of the properties and environments of hard X-ray selected AGN, which have shown that they are associated with an excess of galaxy mergers \citep[][see also \citealt{2018ApJ...858..110P}]{2010ApJ...716L.125K, 2011ApJ...739...57K, 2012ApJ...746L..22K, 2018Natur.563..214K}. However, while low-$z$ galaxies selected to host hard X-ray AGN show an enhancement in merger fraction, to date no study has approached the question from the other way around: do low-$z$ galaxy mergers show an enhancement in X-ray AGN? Required for such a study is a large sample of galaxies observed in X-rays, with sufficient depth such that non-detections carry physical significance, and preferably observed serendipitously to avoid a selection bias for galaxies believed to host an AGN \textit{a priori}.

Since its launch in 1999, the \textit{XMM-Newton} telescope has completed over 14\,000 observations. Its relatively poor angular resolution (several arcsec) is compensated by its large field of view (radius~$\sim15\arcmin$) and sensitivity in the 0.2--10~keV band comparable to the \textit{Chandra X-ray Observatory}. Owing to this large field of view, the vast majority of sources found in \textit{XMM-Newton} data are serendipitous, and the latest iteration of the XMM source catalog \citep[4XMM~DR9;][]{2019A&A...submitted}\footnote{\url{http://xmmssc.irap.omp.eu/Catalogue/4XMM-DR9/4XMM_DR9.html}} has 810\,795 detections of 550\,124 unique sources. While 4XMM is therefore an invaluable resource for large statistical X-ray studies of the properties of \emph{detected} sources, sources observed by \textit{XMM-Newton} that are not in the 4XMM catalog cannot be treated as undetected at some flux limit, because the field of view of \textit{XMM-Newton} is divided into multiple CCDs, separated by significant gaps, each with its own sensitivity and bad pixels. As a result, population studies requiring accurate knowledge of whether or not an \textit{XMM-Newton} observation legitimately constrains the X-ray flux of an object requires an independent analysis of the data, performed in a homogeneous manner for all objects.

Additionally, because the intrinsic X-ray luminosities of AGN are strongly correlated with their mid-IR luminosities \citep[e.g.][]{2015ApJ...798...38S, 2015ApJ...807..129S,2015MNRAS.454..766A}, the ratio of the \emph{apparent} hard X-ray (2--10~keV) luminosity of an AGN, which \textit{XMM-Newton} can measure, and its mid-IR luminosity may provide insight into the line-of-sight absorption to the AGN \citep{2017ApJ...848..126S}, which is predicted to reach its peak in late-stage galaxy mergers when AGN activity is also more likely to be selected in the mid-IR \citep{2018MNRAS.478.3056B}. Finally, while mid-IR selection is insensitive at lower AGN luminosities (relative to its host galaxy), hard X-ray selection remains highly reliable down to luminosities of $L_\mathrm{2-10~keV}\sim10^{40}$~erg~s$^{-1}$, below which X-ray binary activity in extreme starbursts may be a contaminant in lower angular resolution data \citep[e.g.][]{2018ApJ...865...43F, 2019ApJS..243....3L}

In this work, we carry out a detailed assessment of both X-ray detections and non-detections in a sample of 43 post-merger systems, the merger stage that exhibits the greatest AGN enhancement in the optical and mid-IR \citep[e.g.][]{2013MNRAS.435.3627E,2014MNRAS.441.1297S}, which we then compare to results obtained at mid-IR and visual (optical) wavelengths. Throughout this work, we use a flat $\Lambda$CDM cosmology with $H_0=70$~km~s$^{-1}$~Mpc$^{-1}$ and $\Omega_\mathrm{M}=0.3$. Unless otherwise noted, all X-ray fluxes quoted in this work are observed-frame 2--10~keV and all X-ray luminosities are rest-frame 2--10~keV and apparent (not corrected for intrinsic $N_\mathrm{H}$). All stellar masses and star formation rates (SFRs) are given as $\log_{10}(M_{\star} / M_{\sun})$ and $\log_{10}(M_{\star} / M_{\sun}~\mathrm{yr}^{-1})$.

\section{Methods}
\subsection{Post-merger and Control Selection} \label{sec: Postmerger and Control Selection}
Our post-merger sample is assembled from two components: 1) new XMM observations of previously identified PMs from the SDSS; and 2) a visual search for PMs in existing XMM archival observations.  For the first component, we selected post-mergers from a larger, visually-classified sample of galaxy pairs and mergers selected from the SDSS, Data Release 7 \citep[DR7;][]{2009ApJS..182..543A}, and described in detail in \citet{2013MNRAS.435.3627E}, Sections 2.2 and 2.3. Briefly, a galaxy is considered to be a post-merger if it shows strong morphological asymmetry, tidal tails/streams, the outer remnants of another galaxy, or all of the above, and does not have a close companion indicating an ongoing merger or flyby. We limit our sample to those with redshifts between 0.01 and 0.2, and we use stellar masses and SFRs from the MPA-JHU catalog for DR7,\footnote{\url{https://wwwmpa.mpa-garching.mpg.de/SDSS/DR7}} with stellar masses following \citet{2003MNRAS.341...33K,2007ApJS..173..267S} and SFRs following \citet{2004MNRAS.351.1151B}. We acquired \textit{XMM-Newton} observations for 13 post-mergers during cycle AO15 (proposal ID 078515; PI: Ellison). These observations were designed to detect a heavily-absorbed ($N_\mathrm{H} = 5 \times 10^{23}$~cm$^{-2}$) AGN with an intrinsic 2--10~keV X-ray luminosity of $10^{42}$~erg~s$^{-1}$ with 50 counts between 0.3--10~keV. 

To complement our new observations, we searched for post-mergers with archival \textit{XMM-Newton} observations by cross-matching all galaxies with stellar mass measurements from the MPA-JHU catalog to the \textit{XMM-Newton} observation log from 24~November~2019 to within $15\arcmin$, the approximate radius of the EPIC field of view, returning 36\,552 objects with 72\,635 science observations. We then matched these objects onto the Galaxy~Zoo for SDSS DR7 \citep{2008MNRAS.389.1179L,2011MNRAS.410..166L},\footnote{\url{https://data.galaxyzoo.org}}, and visually inspected the SDSS thumbnails for all objects with a merger vote fraction of \texttt{P\_MG}~$\geq0.3$ to identify additional post-mergers.

We selected control galaxies in the following manner. For each post-merger, we matched all galaxies not in our post-merger list to within $\pm0.1$~dex in stellar mass, $\pm0.01$ in redshift, and $\pm0.1$~dex in normalised environment parameter $\delta_5$, as described in \citet[][Section 2.2]{2013MNRAS.435.3627E}. We manually inspected all candidate control galaxies for each post-merger, further flagging interactions and occasionally finding additional post-mergers. After inspection, we re-ran the control matching procedure, excluding interactions and folding in additional post-mergers, and repeated this process until all candidate controls were inspected and no additional post-mergers were found.

After downloading and reprocessing the data through the pipeline described in Section~\ref{sec: XMM-Newton Pipeline}, we allowed only controls with exposure times equal to or \emph{longer} than their corresponding post-merger. We then reduced the \emph{effective} exposure time of the controls to match that of the post-merger by multiplying their X-ray signal-to-noise by $\sqrt{t_\mathrm{post-merger}/t_\mathrm{control}}$, where $t$ is the exposure time. This ensures that differences in detection fraction are not biased by differences in exposure time, and allows for a much larger number of control galaxies for a given post-merger than would have been possible by trying to find controls with \textit{XMM-Newton} observations matched closely in exposure time. It also has the attractive property that the effective exposure time of a given control is matched \emph{exactly} to the exposure time of its post-merger, and the distribution of exposure times between the sample of post-mergers and their controls is also matched exactly if the number of controls is the same for each post-merger. We enforced a specific number of controls per post-merger by discarding post-mergers with fewer than the required number and removing controls with the largest redshift offsets for post-mergers with greater than the required number. Our final sample consists of 43 post-mergers, each with 10 controls (Figures~\ref{fig:pm}, \ref{fig:mzs_hist}, \ref{fig:m_vs_z}; Table~\ref{tab:stats}).

\begin{figure*}
    \includegraphics[width=\textwidth]{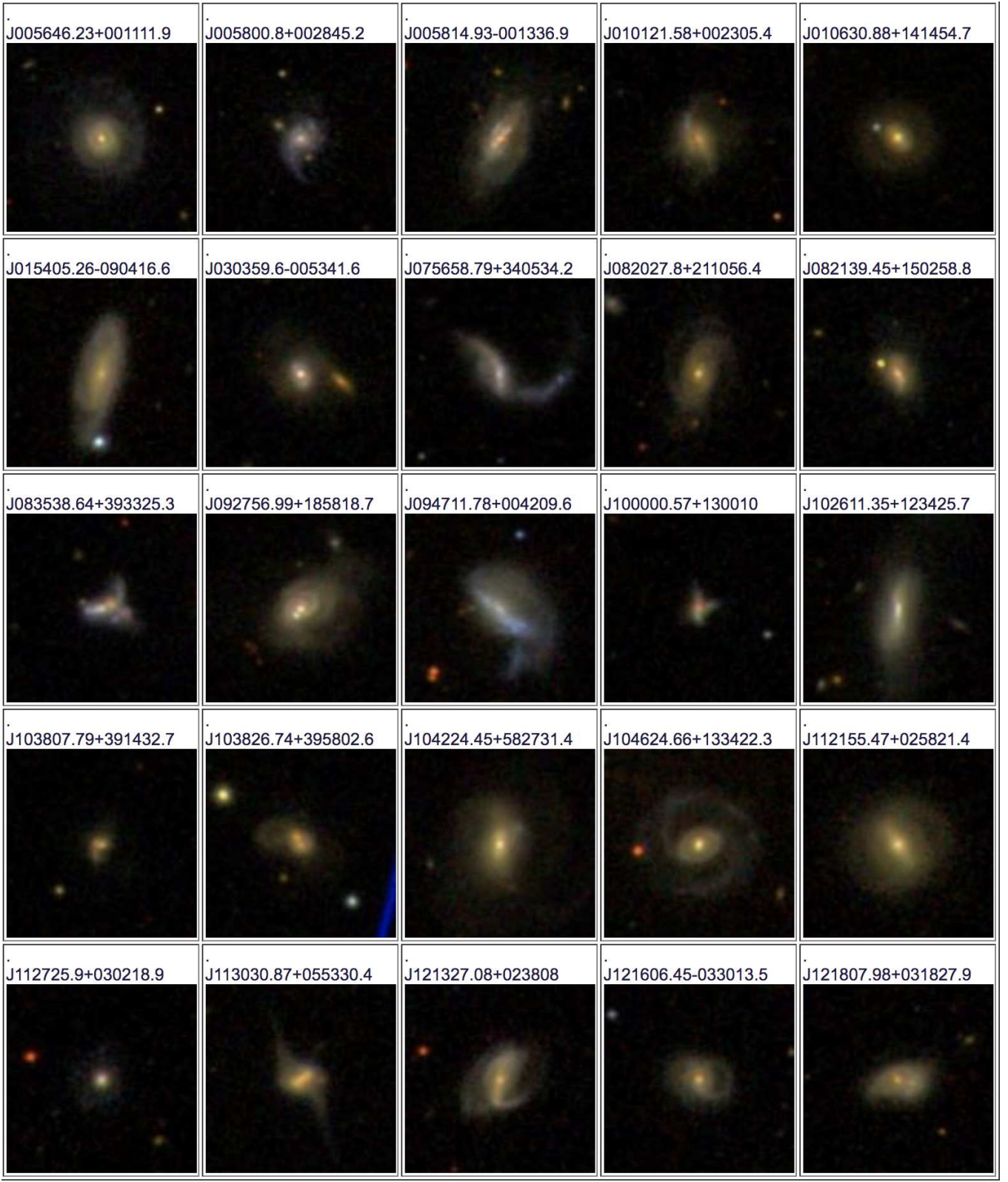}
        \caption{SDSS DR16 $50\arcsec\times50\arcsec$ thumbnails of the 43 post-mergers studied in this work, ordered by increasing right ascension. Objects detected in X-rays have the logarithm of their rest-frame, apparent X-ray luminosities, in erg~s$^{-1}$, listed above their SDSS designations.}
    \label{fig:pm}
\end{figure*}

\begin{figure*}
    \includegraphics[width=\textwidth]{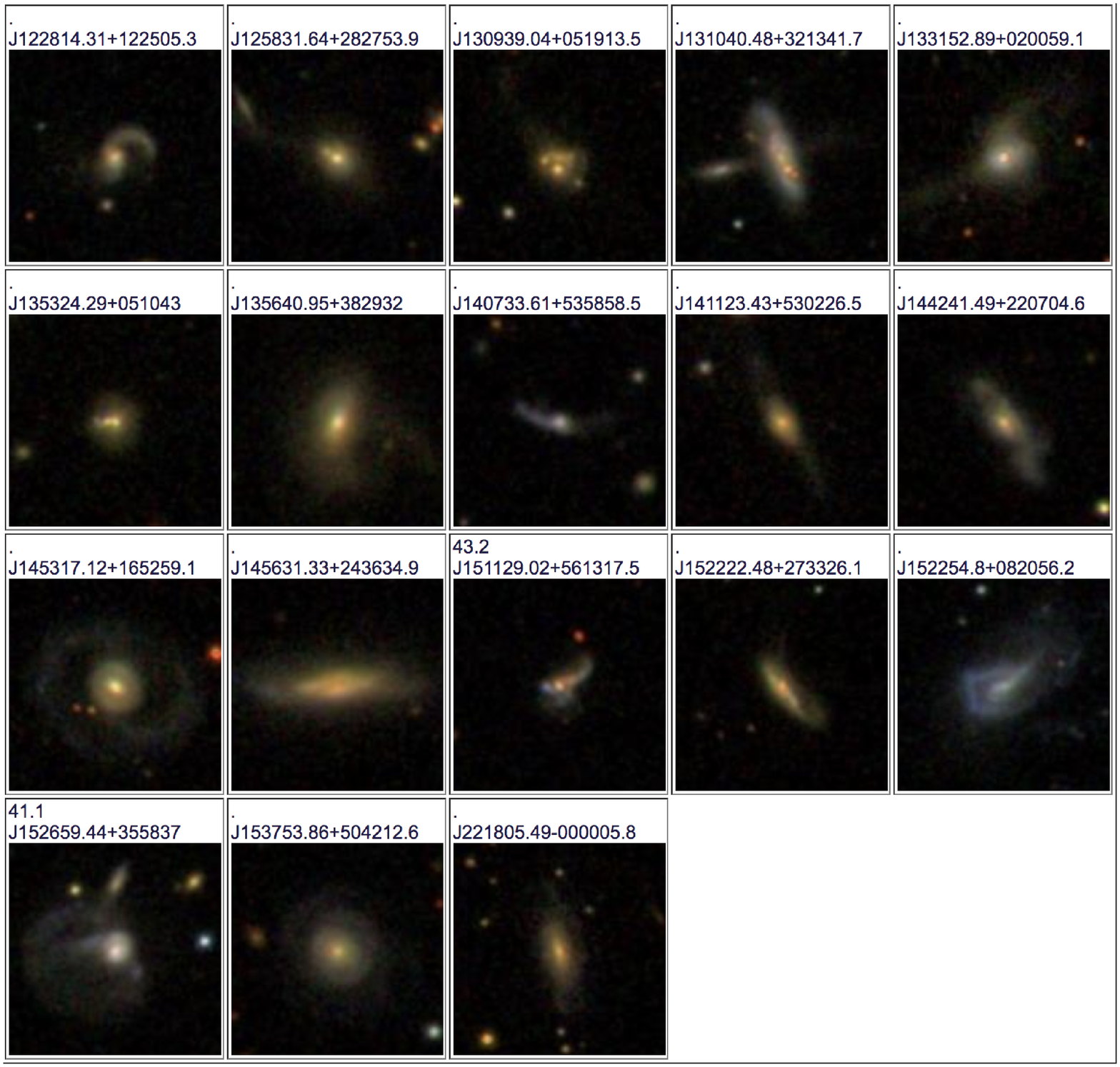}
        \contcaption{}
\end{figure*}

\begin{figure*}
    \includegraphics[width=1.0\textwidth]{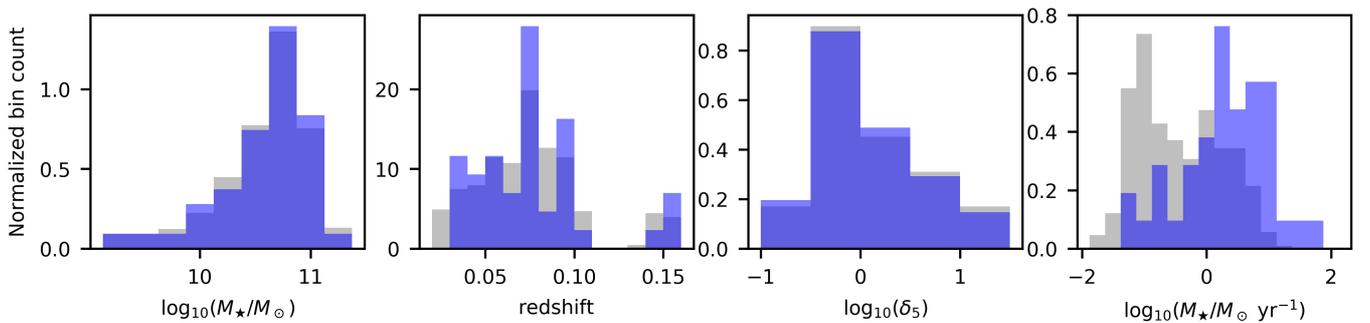}
        \caption{Distributions of post-mergers (blue) and their matched controls (grey), normalized such that the total area under the histogram for each population sums to 1. Note that the controls have not been matched in SFR.}
    \label{fig:mzs_hist}
\end{figure*}

\begin{figure}
    \includegraphics[width=\columnwidth]{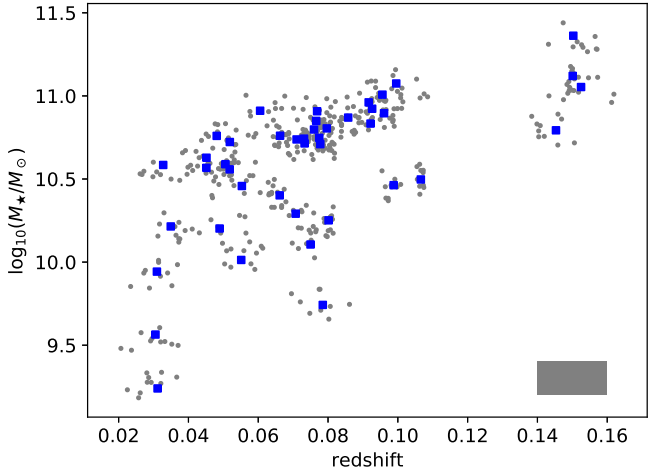}
        \caption{Post-mergers (blue filled squares) with their matched controls (grey filled circles). There are exactly 10 controls per post-merger. The gray rectangle shows the matching tolerance in redshift and stellar mass around each post-merger.}
    \label{fig:m_vs_z}
\end{figure}

\begin{table*}
    \centering
    \caption{Mean values (and standard errors) of the properties of the post-mergers and their matched control sample, with KS test $p$-value.}
    \label{tab:stats}
    \begin{tabular}{lrrrrrr}
        \hline
        Property & Post-merger ($N=43$) & Control ($N=430$) & $p_\mathrm{KS}$ \\
        \hline
        $\log_{10}(M_{\star} / M_{\sun})$                             & $10.59\pm0.06$   & $10.58\pm0.02$ & 0.95 \\
        redshift                                                                    & $0.076\pm0.005$ & $0.076\pm0.002$ & 0.97 \\
        $\log_{10}(\delta_5)$                                           & $0.15\pm0.10$    & $0.15\pm0.03$      & 1.00 \\
        $\log_{10}(M_{\star} / M_{\sun}~\mathrm{yr}^{-1})$ & $0.34\pm0.11$     & $-0.44\pm0.03$ & $2.5\times10^{-7}$ \\
        \hline
    \end{tabular}
\end{table*}

\subsection{XMM-Newton Pipeline} \label{sec: XMM-Newton Pipeline}
We developed a pipeline that downloads, reprocesses, filters, and extracts source counts and upper limits for \textit{XMM-Newton} data. This carries the advantage that all source counts and upper limits are extracted from the same aperture, using data reprocessed with the same Current Calibration Files (CCFs), and so produces fluxes that are directly comparable between the archival data, including the observations from AO15. Moreover, it also allows us to say whether or not a source is detected. The pipeline iterates over directories corresponding to each ObsID that contain a set of valid Observation Data Files (ODFs). For each ODF, the Science Analysis System (SAS),\footnote{\url{https://www.cosmos.esa.int/web/xmm-newton/sas}} version 18.0.0, routine \texttt{cifbuild} is called to produce a CCF pertaining to the observation. A SAS summary file is produced using \texttt{odfingest}, and then the pn/MOS event files are produced using \texttt{epchain}/\texttt{emchain}. For observations containing both scheduled and unscheduled exposures, single pn/MOS event files are produced using \texttt{evlistcomb}. The pn and MOS event files are then filtered for science-grade events using \texttt{evselect} with the canned screening expressions \texttt{\#XMMEA\_EP}/\texttt{\#XMMEA\_EM}, and requiring \texttt{PATTERN<4} for pn and \texttt{PATTERN<12} for MOS. We made high energy background light curves with \texttt{(PI>10000\&\&PI<12000)} for pn and \texttt{(PI>10000\&\&PI<15000)} for MOS, and we removed periods in which the total count rate exceeded 0.35~cps for pn and 0.40~cps for MOS.\footnote{\url{https://www.cosmos.esa.int/web/xmm-newton/sas-thread-epic-filterbackground}} 2--10~keV X-ray images are produced using \texttt{evselect} from which source counts are extracted using \texttt{eregionanalyse}. A mean background value is supplied to \texttt{eregionanalyse} derived from a source-masked version of the input image, and off-axis vignetting is accounted for by supplying \texttt{eregionanalyse} with an exposure map generated using \texttt{eexpmap}. The exposure map is also used to ensure that the source region does not fall within a chip gap or other area with censored coverage.

\subsubsection{Local Background} \label{subsubsec: Local Background}
A small fraction of extraction apertures ($\sim9.4\%$) are within a region of high \emph{local} background, usually either due to the presence of diffuse emission or a neighboring very bright source. We effectively removed these by flagging all sources with a background rate treater than $>1\times10^{-5}$~count~arcsec$^{-2}$~s$^{-1}$ for pn and greater than $>5\times10^{-6}$~count~arcsec$^{-2}$~s$^{-1}$ for MOS. A smaller fraction ($\sim1.1\%$) of source extractions have an anomalously \emph{low} local background that we found in some cases to be due to \texttt{eexpmap} producing an exposure map with CCDs considered to be active when no events were recorded, which can happen when the attitude changes during individual exposures. We effectively removed anomalously low-background sources by requiring a background rate greater than $2\times10^{-7}$ count~arcsec$^{-2}$~s$^{-1}$ for all data. After making these background cuts, the distributions of local backgrounds are approximately log-normally distributed with means $-5.48$ for pn and $-6.01$ for MOS.

\subsubsection{Uncertainty Correction} \label{subsubsec: Uncertainty Correction}
The mean background value is supplied to \texttt{eregionanalyse} as a constant, so the calculated formal error does not account for the uncertainty of the background. To correct this, we added in quadrature to the formal count rate errors $\pi R_\mathrm{src}^2 \sigma_\mathrm{bkg} N^{-1/2} t^{-1} \mathrm{EEF}^{-1}$, where $R_\mathrm{src}$ is the source extraction aperture in arcsec, $\sigma_\mathrm{bkg}$ is the background standard deviation in count~arcsec$^{-2}$, $N$ is the number of arcsec$^2$ in the source region, $\mathrm{EEF}$ is the enclosed energy fraction corresponding to $R_\mathrm{src}$, and $t$ is the exposure time at the position of the source.

\subsubsection{X-ray Fluxes} \label{subsubsec: X-ray Fluxes}
We converted the per-source, per-observation count rates to rest-frame fluxes by using energy conversion factors (ECFs) we obtained using the Portable, Interactive Multi-Mission Simulator (PIMMs), version 4.10, via WebPIMMS.\footnote{\url{https://heasarc.gsfc.nasa.gov/cgi-bin/Tools/w3pimms/w3pimms.pl}} We calculated the redshift-dependent (K corrected), filter-dependent ECF for each source using a lookup table, given in Table~\ref{tab:ecfs}, and linearly interpolating between the redshifts. The per-source, per-observation fluxes were then converted to per-source fluxes across all observations by adding the total energy per cm$^{2}$ across all observations and dividing by the total exposure time:

\begin{equation}
F, \sigma_F = \frac{\Sigma F_i t_i}{\Sigma t_i}, \frac{\sqrt{\Sigma (\sigma_{F_i }t_i)^2}}{\Sigma t_i}
\end{equation}

\begin{table}
    \centering
    \caption{ECFS used in this work, obtained using PIMMS, version 4.10, converting between observed 2--10~keV count rates and rest-frame 2--10~keV flux uncorrected for intrinsic absorption. ECFs ($10^{11}$~count~cm$^2$~erg$^{-1}$) assume a power-law spectrum with $\Gamma=1.7$, Galactic $N_\mathrm{H}=3\times10^{20}$~cm$^{-2}$, and no intrinsic absorption.}
    \label{tab:ecfs}
    \begin{tabular}{ccccc}
        \hline
        Camera & redshift & Thin & Medium & Thick \\
        \hline
	pn & 0.00 & 1.224 & 1.263 & 1.213 \\
	pn & 0.05 & 1.242 & 1.282 & 1.231 \\
	pn & 0.10 & 1.260 & 1.300 & 1.248 \\
	pn & 0.15 & 1.276 & 1.317 & 1.265 \\
	pn & 0.20 & 1.293 & 1.335 & 1.282 \\
	MOS & 0.00 & 0.436 & 0.439 & 0.423 \\
	MOS & 0.05 & 0.443 & 0.446 & 0.430 \\
	MOS & 0.10 & 0.449 & 0.452 & 0.436 \\
	MOS & 0.15 & 0.455 & 0.458 & 0.442 \\
	MOS & 0.20 & 0.461 & 0.464 & 0.447 \\
        \hline
    \end{tabular}
\end{table}

\noindent After filtering for the local background (Section~\ref{subsubsec: Local Background}), we found that the distribution of fluxes divided by their errors (S/N) follows a normal distribution centred around zero with a sigma of unity for offsets less than 3. We therefore consider detections above $3\sigma$ (99.9\% for a normal distribution) to be a natural demarcation between non-detections and detections, which we employ in this work. 

\subsection{Small Number Statistics} \label{subsec: Small Number Statistics}
We use binomial statistics \citep[e.g.][]{1986ApJ...303..336G} to account for uncertainties inherent to estimating the fraction $n/N$ of a population with some property when limited to a small sample size $N$ (i.e.\ the fraction of post-mergers with AGN when the total number of post-mergers is small). When calculating the null hypothesis $p$-value of $n$ AGN being observed in a sample of size $N$ given an expected frequency $f$, we use \texttt{numpy.random.binomial} and quote the probability of $n$ or a larger number being observed if $n/N$ is larger than $f$, and a probability of $n$ or smaller being observed if $n/N$ is smaller than $f$. Under the null hypothesis that the frequency of AGN in post-mergers is the same as that seen in their controls, for example, the best estimate for the expected frequency $f$ is $(n_\mathrm{post-merger} + n_\mathrm{control}) / (N_\mathrm{post-merger} + N_\mathrm{control})$. To compare the excess of AGN in post-mergers compared with their controls, defined as $(n_\mathrm{post-merger} / N_\mathrm{post-merger}) / (n_\mathrm{control} / N_\mathrm{control})$, we generate $10^6$ random samples of each quantity drawn from the binomial distribution using \texttt{numpy.random.binomial} and divide these samples, returning the values at percentile $(100\pm\mathrm{CL})/2$, where CL is the confidence limit. In the case where $n_\mathrm{control}=0$, we use the upper CL bound on $n_\mathrm{control}/N_\mathrm{control}$ and calculate the corresponding limit on $(n_\mathrm{post-merger} / N_\mathrm{post-merger}) / (n_\mathrm{control} / N_\mathrm{control})$. Throughout this work, the confidence interval is 90\%, so upper and lower limits may be interpreted as 95\%, 5\%.

\section{Results and Discussion} \label{sec: Results and Discussion}
\subsection{X-rays} \label{subsec: X-rays}
Two of the 43 post-mergers are detected in X-rays ($4.7^{+9.3}_{-3.8}\%$) above the adopted detection threshold of $3\sigma$ (Table~\ref{tab:pm}), both of which have rest-frame, apparent X-ray luminosities greater than $10^{41}$~erg~s$^{-1}$. Nine of the 430 control galaxies are detected  ($2.1^{+1.5}_{-1.0}\%$), corresponding to an X-ray detection excess in the post-mergers of $2.22^{+4.44}_{-2.22}$. While this excess is not statistically significant ($p=0.26$), it is consistent with the factor of 3.75 optical AGN excess found in \citet{2013MNRAS.435.3627E}, but potentially consistent with no excess at all. The 95\% upper limit on the X-ray excess of 6.7 is, however, inconsistent with the factor of 11--20 mid-IR AGN excess in post-mergers found by \cite{2014MNRAS.441.1297S}, suggesting that the X-ray counterparts of the mid-IR AGNs are not being detected.

\begin{table*}
\centering
\caption{Properties of the 43 post-mergers examined in this work, with their total \textit{XMM-Newton} exposure time, X-ray luminosities (observed, rest-frame 2--10~keV luminosity, with $3\sigma$ upper limits for non-detections), AllWISE colour, and optical classification \citep[quiescent, star-forming, or AGN classified following][]{2001ApJ...556..121K, 2003MNRAS.346.1055K, 2006MNRAS.371..972S}. Note that any object meeting the \citet{2001ApJ...556..121K} AGN criterion also meets the less strict \citet{2003MNRAS.346.1055K} criterion, which also meets the even less strict \citet{2006MNRAS.371..972S} criterion.}
\label{tab:pm}
\begin{tabular}{crrrrrrrrr}
\hline
           & R.A. & Decl. &              &                                                                               &                                                                                                       & exptime &                                                                                                       & $W1-W2$ &                   \\ [-0.25cm]
Name &          &         & redshift  & $\log_{10}\left(\frac{M_{\star}}{M_{\sun}}\right)$   & $\log_{10}\left(\frac{M_{\star}}{M_{\sun}}\mathrm{yr}^{-1}\right)$ &               & $\log_{10}\left(\frac{L_\mathrm{2-10~keV}}{\mathrm{erg~s^{-1}}}\right)$  &                  & Opt.\ class \\ [-0.25cm]
           & deg  & deg   &              &                                                                               &                                                                                                        & kilosec &                                                                                                        & mag          &                   \\
\hline
J0056$+$0011 & 14.19265 & 0.18665 & 0.05182 & 10.6 & $-$0.06 & 8.7 & <41.1 & 0.159 & Q \\
J0058$+$0028 & 14.50334 & 0.47922 & 0.08016 & 10.3 & 0.75 & 7.1 & <41.8 & 0.259 & K03 \\
J0058$-$0013 & 14.56222 & $-$0.22694 & 0.07099 & 10.7 & 1.11 & 6.0 & <41.5 & 0.336 & S06 \\
J0101$+$0023 & 15.33994 & 0.38485 & 0.09264 & 10.9 & 0.49 & 14.8 & <41.4 & 0.301 & K03 \\
J0106$+$1414 & 16.62870 & 14.24853 & 0.07598 & 10.8 & $-$1.21 & 28.6 & <41.4 & 0.080 & Q \\
J0154$-$0904 & 28.52194 & $-$9.07129 & 0.05061 & 10.6 & $-$0.03 & 1.7 & <41.9 & 0.043 & Q \\
J0303$-$0053 & 45.99835 & $-$0.89491 & 0.06615 & 10.4 & 0.79 & 14.8 & <41.4 & 0.210 & S06 \\
J0756$+$3405 & 119.24497 & 34.09284 & 0.07073 & 10.3 & 0.67 & 30.6 & <41.3 & 0.273 & SF \\
J0820$+$2110 & 125.11584 & 21.18233 & 0.09954 & 11.1 & $-$0.05 & 50.0 & <41.3 & 0.101 & Q \\
J0821$+$1502 & 125.41440 & 15.04969 & 0.06626 & 10.8 & 0.96 & 3.2 & <42.0 & 0.314 & SF \\
J0835$+$3933 & 128.91100 & 39.55703 & 0.09880 & 10.5 & 0.88 & 4.5 & <41.9 & 0.452 & S06 \\
J0927$+$1858 & 141.98749 & 18.97188 & 0.05186 & 10.7 & 0.65 & 9.6 & <41.1 & 0.163 & S06 \\
J0947$+$0042 & 146.79909 & 0.70269 & 0.03055 & 9.6 & 0.52 & 2.5 & <41.2 & 0.254 & SF \\
J1000$+$1300 & 150.00241 & 13.00278 & 0.05512 & 10.0 & 0.29 & 22.2 & <41.6 & 0.340 & K03 \\
J1026$+$1234 & 156.54729 & 12.57381 & 0.03096 & 9.9 & 0.13 & 5.5 & <40.8 & 0.066 & SF \\
J1038$+$3914 & 159.53247 & 39.24243 & 0.14532 & 10.8 & 0.13 & 18.7 & <42.0 & 0.621 & K01 \\
J1038$+$3958 & 159.61144 & 39.96741 & 0.15028 & 11.4 & 1.09 & 5.5 & <42.5 & $\cdots$ & K03 \\
J1042$+$5827 & 160.60191 & 58.45875 & 0.04517 & 10.6 & $-$1.23 & 4.4 & <41.0 & 0.009 & K01 \\
J1046$+$1334 & 161.60277 & 13.57286 & 0.09165 & 11.0 & 0.22 & 15.6 & <41.6 & 0.157 & K03 \\
J1121$+$0258 & 170.48114 & 2.97262 & 0.04810 & 10.8 & $-$0.63 & 5.0 & <41.4 & $-$0.015 & Q \\
J1127$+$0302 & 171.85792 & 3.03859 & 0.07498 & 10.1 & 0.15 & 19.4 & <41.5 & 0.081 & S06 \\
J1130$+$0553 & 172.62864 & 5.89178 & 0.03493 & 10.2 & $-$0.22 & 14.1 & <40.6 & 0.306 & Q \\
J1213$+$0238 & 183.36287 & 2.63556 & 0.07311 & 10.7 & 0.37 & 16.2 & <41.5 & 0.176 & K03 \\
J1216$-$0330 & 184.02692 & $-$3.50376 & 0.09217 & 10.8 & 0.28 & 12.2 & <41.5 & 0.189 & K01 \\
J1218$+$0318 & 184.53329 & 3.30776 & 0.07777 & 10.7 & 0.54 & 26.5 & <41.2 & 0.185 & Q \\
J1228$+$1225 & 187.05964 & 12.41816 & 0.15250 & 11.1 & 1.39 & 39.7 & <42.0 & 0.469 & K03 \\
J1258$+$2827 & 194.63184 & 28.46499 & 0.09605 & 10.9 & $-$0.45 & 13.5 & <41.6 & 0.198 & K01 \\
J1309$+$0519 & 197.41271 & 5.32043 & 0.10659 & 10.5 & 0.94 & 4.8 & <42.1 & 0.275 & K03 \\
J1310$+$3213 & 197.66868 & 32.22826 & 0.04892 & 10.2 & 0.75 & 20.6 & <41.1 & 0.370 & SF \\
J1331$+$0200 & 202.97039 & 2.01643 & 0.08578 & 10.9 & 1.69 & 13.3 & <41.7 & 1.256 & K01 \\
J1353$+$0510 & 208.35123 & 5.17862 & 0.07749 & 10.7 & $-$0.69 & 0.1 & <42.7 & 0.001 & Q \\
J1356$+$3829 & 209.17065 & 38.49224 & 0.06054 & 10.9 & $-$0.70 & 7.1 & <41.4 & 0.020 & Q \\
J1407$+$5358 & 211.89005 & 53.98294 & 0.07850 & 9.7 & 0.34 & 8.1 & <41.7 & 0.143 & SF \\
J1411$+$5302 & 212.84763 & 53.04072 & 0.07687 & 10.9 & $-$0.15 & 8.2 & <41.4 & 0.245 & K03 \\
J1442$+$2207 & 220.67289 & 22.11795 & 0.07964 & 10.8 & 1.33 & 6.0 & <41.9 & 0.342 & K03 \\
J1453$+$1652 & 223.32134 & 16.88309 & 0.04515 & 10.6 & $-$0.90 & 23.5 & <40.9 & 0.013 & K01 \\
J1456$+$2436 & 224.13055 & 24.60972 & 0.03277 & 10.6 & $-$0.07 & 20.1 & <40.6 & 0.844 & K01 \\
J1511$+$5613 & 227.87092 & 56.22155 & 0.15012 & 11.1 & 2.22 & 9.3 & 43.2 & 0.805 & S06 \\
J1522$+$2733 & 230.59369 & 27.55726 & 0.07327 & 10.7 & 0.48 & 46.1 & <41.3 & 0.291 & K01 \\
J1522$+$0820 & 230.72835 & 8.34895 & 0.03116 & 9.2 & $-$0.24 & 19.6 & <40.5 & 0.303 & SF \\
J1526$+$3558 & 231.74767 & 35.97695 & 0.05531 & 10.5 & 1.00 & 32.9 & 41.1 & 1.497 & K03 \\
J1537$+$5042 & 234.47443 & 50.70353 & 0.07661 & 10.8 & 0.44 & 16.4 & <41.6 & 0.144 & Q \\
J2218$-$0000 & 334.52289 & $-$0.00162 & 0.09559 & 11.0 & 0.77 & 9.4 & <41.9 & 0.189 & Q \\
\hline
\end{tabular}
\end{table*}

We reiterate that we have not controlled for SFR in our assembly of the control sample (Table~\ref{tab:stats}; Figure~\ref{fig:mzs_hist}, right). This was intentional, as we allow for the possibility that the process that fuels star formation may also fuel AGN activity. Indeed, the mean SFR of the post-mergers is 9.0~$M_{\sun}$~yr$^{-1}$, $7.9^{+6.6}_{-4.5}$ times larger than the mean SFR of the controls (Figure~\ref{fig:mzs_hist}), but consistent with with the factor of 3.5 excess seen in the sample of post-mergers studied in \citet{2013MNRAS.435.3627E}, given the uncertainty.\footnote{Uncertainty estimated by bootstrapping the post-mergers and controls.}  While the factor of $\sim2$ excess of X-ray detections in the post-mergers is not significant, we explore the possibility that this number has been spuriously enhanced by greater X-ray binary (XRB) activity in the post-mergers owing to their higher SFRs. We calculated the \emph{expected} X-ray luminosity from X-ray binary (XRB) activity using the Equation~3 from \citet{2010ApJ...724..559L}, which is a function of both stellar mass and SFR, with the best-fit parameters given in their Table~4. The mean expected X-ray luminosity from XRB activity for the detected objects is $1.5\times10^{41}$~erg~s$^{-1}$ for the post-mergers and $1.1\times10^{40}$~erg~s$^{-1}$ for the controls, 1.7 and 2.5 dex lower than the mean observed X-ray luminosities for both populations ($7.6\times10^{42}$~erg~s$^{-1}$ and $3.8\times10^{42}$, respectively), and all of the post-mergers and controls are respectively at least 0.82 and 1.6~dex more luminous in observed X-rays than what is expected from XRB activity. Given the 0.34~dex intrinsic scatter of the relation from \citet{2010ApJ...724..559L}, we conclude that XRBs are not a significant contaminant in the observed X-ray luminosities of our sample.

\subsection{Mid-Infrared} \label{subsec: Mid-Infrared}
To compare our X-ray based results to other multi-wavelength AGN metrics, we matched the 43 post-mergers and their 430 controls to the AllWISE catalog within the default $10\arcsec$ tolerance. For matches with offsets greater than $2\arcsec$, we manually checked the AllWISE counterpart positions, removing any spurious matches (e.g.\ because of the presence of a neighbouring foreground star), leaving 42 out of 43 post-mergers and all 430 controls with AllWISE counterparts. The loss of one post-merger did not significantly affect how well-matched the remaining post-mergers are to the controls, having stellar mass, redshift, and environment KS test $p$ values of 0.99, 0.88, and 1, respectively. The WISE colour $W1-W2$ ($[3.4\micron]-[4.6\micron]$), listed for the post-mergers in Table~\ref{tab:pm}, is especially sensitive to the presence of an AGN, nearly independent of extinction \citep[e.g,][]{2012ApJ...748..142D}, and increases with increasing AGN dominance. We find that the mean $W1-W2$ colour is significantly redder for the post-mergers ($0.297\pm0.047$~mag) than for the controls ($0.138\pm0.005$~mag). While the elevated $W1-W2$ seen in the post-mergers is consistent with a greater frequency and/or dominance of AGN activity, to further quantify this we determined the number of post-mergers and controls with $W1 - W2 > 0.5$, when \emph{some} amount of AGN contribution is generally required \citep[e.g.][]{2018ApJ...858...38S} even at high specific star formation rates \citep{2018MNRAS.478.3056B}, $W1 - W2 > 0.8$, the threshold above which the object is dominated by an AGN, and $W1-W2>1.0$, at which point the object's mid-IR emission is almost purely AGN \citep[e.g.][]{2010ApJ...713..970A, 2012ApJ...753...30S}. We find (Table~\ref{tab:allwise}) that, given the uncertainties, the post-mergers exhibit fractions of objects with $W1-W2$ greater than 0.5, 0.8 consistent with the post-mergers studied in \citet{2014MNRAS.441.1297S}, which exhibited fractions of 0.16 and 0.1, respectively.

\begin{table}
    \centering
    \caption{Number of post-mergers and controls above a given $W1-W2$. The fraction $n/N$ of the total (42 post-mergers and 430 controls; see Section~\ref{subsec: Mid-Infrared}) is given as a percentage along with the null probability that the number of post-mergers above the $W1-W2$ threshold follows the frequency seen in the controls. Uncertainty bounds are $\pm95\%$. For the case where $n_\mathrm{C}=0$, the excess corresponds to the lower limit on $(\%)_\mathrm{P}$ and the upper limit on $(\%)_\mathrm{C}$ that gives a joint confidence limit of 95\%.}
    \label{tab:allwise}
    \begin{tabular}{crrrrrr}
        \hline
        $W1-W2$ & $n_\mathrm{P}$ & $n_\mathrm{C}$ & $(\%)_\mathrm{P}$ & $(\%)_\mathrm{C}$ & $p$ & Excess \\
        \hline
        0.5 & 5 & 3 & $12^{+12}_{-7}$ & $0.7^{+1.1}_{-0.5}$ & $0.00071$ & $17^{+96}_{-12}$ \\ [0.1cm]
        0.8 & 4 & 1 & $10^{+11}_{-6}$ & $0.2^{+0.9}_{-0.2}$ & $0.0010$ &  $41^{+\infty}_{-31}$ \\ [0.1cm]
        1.0 & 2 & 0 & $4.8^{+9.5}_{-3.9}$ & $0.0^{+0.7}_{-0.0}$ & $0.014$ & $\gtrsim6.1$ \\ [0.1cm]
        \hline
    \end{tabular}
\end{table}

We tested the null hypothesis that the post-mergers and their controls exhibit the same frequencies of X-ray and mid-IR AGN, given the observed excesses. To do this, we generated $10^7$ samples from four binomial distributions, two for the X-ray AGN and two for the mid-IR AGN in the post-mergers and controls, using null hypothesis frequencies as defined in Section~\ref{subsec: Small Number Statistics}. We divided the post-merger samples by the control samples to generate the X-ray and mid-IR AGN excesses. Divide-by-zero instances were set to $+\infty$, and $0/0$ instances were discarded. We then divided the X-ray AGN excesses by the mid-IR AGN excesses and counted the number of instances where the X-ray excess was less than the observed value of 2.2 \emph{and} the mid-IR excess was above the observed value of 17. We find that the probability under the null hypothesis of getting the observed X-ray excess and the observed mid-IR AGN excess is $2.7\times10^{-4}$ for $W1-W2>0.5$ and $3.2\times10^{-3}$ for $W1-W2>0.8$. Post-mergers therefore exhibit a significant decrement in the frequency of X-ray AGN, given the excess of mid-IR AGN they show over the controls. Put  differently, the non-merger control galaxies have $\sim8$ times more X-ray AGN per AGN with $W1-W2>0.5$ and $\sim18$ times more X-ray AGN per AGN with $W1-W2>0.8$ than the post-mergers. Given the tight relationship between the X-ray and mid-IR luminosities of AGN \citep[e.g.][]{2015MNRAS.454..766A}, this suggests that the observed X-ray luminosities in the post-mergers are significantly attenuated by line-of-sight absorption, a result predicted by numerical simulations \citep{2018MNRAS.478.3056B}, and supported by other empirical work \citep[e.g.][]{2017MNRAS.468.1273R,2018ApJ...853...63D,2018PASJ...70S..37G}. To explore this, we calculated the ratio of the 2--10~keV to $12\micron$ monochromatic luminosities, which can be used to estimate column density $N_\mathrm{H}$ \citep[e.g.][]{2017ApJ...848..126S}, for the mid-IR AGN with $W1-W2>0.5$. We converted their $W1-W2$ colours to spectral indices \citep[e.g.][Table 1]{2010AJ....140.1868W} and calculated their rest-frame $L_{12\micron}$. We converted $L_{12\micron}$ to predicted $L_\mathrm{2-10~keV}$ using the relation for unabsorbed ($N_\mathrm{H}<10^{22}$~cm$^{-2}$) AGN shown in Figure~11 of \citet{2017ApJ...848..126S}:

\begin{equation}
\log_{10}\left(\frac{L_\mathrm{2-10~keV}}{\mathrm{erg~s^{-1}}}\right) = 0.956333 \cdot \log_{10}\left(\frac{L_\mathrm{12\micron}}{\mathrm{erg~s^{-1}}}\right) + 1.60567
\end{equation}

\noindent (C.~Ricci, private communication). For both the post-mergers and the controls, the predicted X-ray luminosities are above the detection threshold, suggesting that the non-detections are physically significant. We set the X-ray luminosity of the 3 non-detections in the post-mergers and the 1 non-detection in the controls to the $3\sigma$ upper limit and calculated the corresponding upper limits on $L_\mathrm{2-10~keV}$/$L_{12\micron}$, finding mean values of $0.07\pm0.04$ for the post-mergers and $0.12\pm0.05$ for the controls ($p=0.46$). Given the low number of objects with $W1-W2>0.5$, we therefore cannot distinguish between the post-mergers and the controls. However, under the null hypothesis that there is no difference in $L_\mathrm{2-10~keV}$/$L_{12\micron}$ between the post-mergers and the controls, the mean value of $L_\mathrm{2-10~keV}$/$L_{12\micron}$ is $\sim0.09\pm0.03$, indicating heavy absorption \citet[][Figure~11]{2017ApJ...848..126S} in mid-IR AGN in general. The higher frequency of mid-IR AGN in post-mergers would therefore correspond to a higher frequency of heavily-absorbed AGN, consistent with the disparity between the X-ray and mid-IR AGN excess seen in the post-mergers.

\subsection{Optical} \label{subsec: Optical}
\begin{figure}
    \includegraphics[width=\columnwidth]{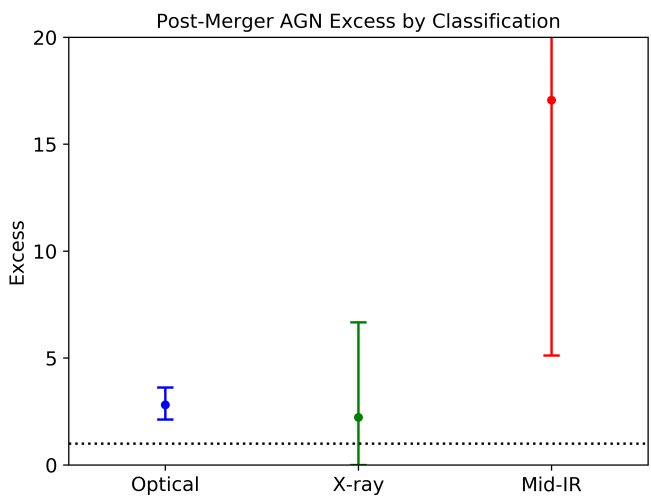}
        \caption{The excess of AGN in post-mergers over their controls, depending on AGN classification, with error bars containing the 90\% confidence interval. The dotted line is at an excess of 1, indicating no difference between the post-mergers and the controls. Optical AGN (meeting the \citealt{2006MNRAS.371..972S} emission line ratio criteria) exhibit a significantly smaller excess than mid-IR AGN ($W1-W2>0.5$~mag), in agreement with previous work; however, the X-ray excess is not significant, with an upper limit of 6.7.}
    \label{fig:agnclassbar}
\end{figure}

Finally, we also explored the optical emission line properties of our sample. Using the MPA-JHU catalog,\footnote{\url{https://www.sdss.org/dr16/spectro/galaxy_mpajhu}} we categorise all objects into those with emission lines dominated by AGN activity using the \citet{2001ApJ...556..121K} demarcation, and objects with decreasing AGN contributions using the \citet{2003MNRAS.346.1055K} and \citet{2006MNRAS.371..972S} demarcations, respectively. We require S/N greater than 5 in H$\beta$, [O\,\textsc{iii}]\,$\lambda5007$, H$\alpha$, and [N\,\textsc{ii}]\,$\lambda6584$ to ensure accurate classification. We consider any objects meeting this S/N threshold but not falling into any AGN classification to be star-forming (SF), and any objects not meeting the S/N threshold to be quiescent. As with the X-ray and mid-IR metrics, we provide a table of the fraction of objects in these categories in Table~\ref{tab:optical}. The large majority of mergers are strong emission line systems compared to the controls, although both the post-mergers and the controls have a similar fraction of optically star-forming systems ($p=0.39$). The raw fraction of AGN in both the post-mergers and controls is sensitive to the selection method, becoming smaller as the line ratio criteria become stricter.  However, the excess of AGN in post-mergers compared to controls is approximately constant for all three diagnostics, at a value of $\sim3$ ($p\lesssim0.01$), consistent given the uncertainties with the factor of 3.75 found by \citet{2013MNRAS.435.3627E} and \citet{2014MNRAS.441.1297S}, and significantly lower than the mid-IR AGN excess (Figure~\ref{fig:agnclassbar}), in line with previous findings.

A potential explanation for the small excess of optical AGN compared to mid-IR AGN is that the post-mergers have a factor of $\sim8$ higher SFR than the controls, with a mean value of $9.0\pm3.9$~$M_{\sun}$~yr$^{-1}$, versus $1.1\pm0.1$~$M_{\sun}$~yr$^{-1}$ in the controls, so emission line dilution from star formation is precluding optical AGN classification at lower AGN luminosities \citep[e.g.][]{2015ApJ...811...26T}, a result also supported by \citet{2010ApJ...716L.125K}. This explanation is disfavoured in our sample, however, as this would imply that optical AGN in post-mergers should be absent from systems without AGN-dominated \textit{WISE} colour. On the contrary, optical AGN, hereafter defined as those meeting the \citet{2006MNRAS.371..972S} criterion to be consistent with \citet{2011MNRAS.418.2043E,2013MNRAS.435.3627E,2014MNRAS.441.1297S}, are $\sim5$ times more frequent in the post-mergers than mid-IR AGN with $W1-W2>0.5$ (Tables~\ref{tab:optical} and \ref{tab:allwise}), and of the 24 post-mergers with optical emission line ratios meeting the \citet{2006MNRAS.371..972S} criterion 19 of them (79\%) have $W1-W2<0.5$ (Table~\ref{tab:pm}). Indeed, the majority of post-mergers are optical AGN ($58^{+13}_{-14}\%$; Table~\ref{tab:optical}), consistent with mergers playing a key role in AGN triggering. These results are therefore consistent with a picture of galaxy mergers triggering AGN activity, with dominant, luminous AGN found more frequently in the final stages of mergers, in line with previous work \citep[e.g.][]{2018MNRAS.480.3562D}. We note that the high frequency of optical AGN in the post-mergers is not in conflict with the heavy obscuration suggested by our results in Section~\ref{subsec: Mid-Infrared}, as the latter is strictly only true for the obscuring column along the line-of-sight column of the observer. Other, low-obscuration sight lines to the AGN may allow for ionizing photons to escape and produce narrow line regions, which can be hundreds or even thousands of parsec in extent \citep[e.g.][and references therein]{2019MNRAS.489..855C}.

\begin{table}
    \centering
    \caption{Number of post-mergers (out of 43) and controls (out of 430) falling into a given optical emission line classification: quiescent, star-forming, \citet{2006MNRAS.371..972S}, \citet{2003MNRAS.346.1055K}, or \citet{2001ApJ...556..121K}. Uncertainty bounds contain the $90\%$ confidence interval.}
    \label{tab:optical}
    \begin{tabular}{lrrrrrr}
        \hline
        Class & $n_\mathrm{P}$ & $n_\mathrm{C}$ & $(\%)_\mathrm{P}$ & $(\%)_\mathrm{C}$ & $p$ & Excess \\
        \hline
        Q      & 11  & 282 & $26^{+13}_{-11}$  & $66^{+4}_{-4}$         & $1.4\times10^{-6}$      & $0.38^{+0.17}_{-0.16}$ \\ [0.1cm]
        SF    &  7  &  59  & $16^{+12}_{-8}$     & $14^{+3}_{-3}$         &  $0.39$                     & $1.2^{+0.8}_{-0.7}$ \\ [0.1cm]
        S06  & 25 &  89  & $58^{+13}_{-14}$   & $21^{+3}_{-3}$         &  $1.9\times10^{-6}$     & $2.8^{+0.8}_{-0.7}$ \\ [0.1cm]
        K03  & 19 &  69  & $44^{+14}_{-13}$   & $16^{+3}_{-3}$         &  $1.0\times10^{-4}$  & $2.8^{+1.0}_{-0.8}$ \\ [0.1cm]
        K01  &  8  &  26  & $19^{+12}_{-9}$     & $6.0^{+2.2}_{-1.8}$  &  $0.011$                   & $3.1^{+2.4}_{-1.6}$ \\ [0.1cm]
        \hline
    \end{tabular}
\end{table}

\section{Conclusions} \label{sec: Conclusions}
We have performed an \textit{XMM-Newton} X-ray analysis of 43 post-merger systems selected from the SDSS~DR7 along with 430 non-merger control galaxies selected to have statistically consistent redshifts, stellar masses, environments, and identical effective X-ray observation depths. These observations are primarily serendipitous observations from the archive, with additional observations from our AO15 program. We analysed the data in a homogeneous way using a custom \textit{XMM-Newton} data reduction pipeline, producing  2--10 keV rest-frame source flux measurements and upper limits for both detected and undetected objects. We compared our results with the mid-IR (AllWISE) and optical spectroscopic (SDSS/MPA-JHU) catalog data for our sample, finding that our objects are consistent with having been drawn from the same population as studied in our previous work. Our primary findings:

\begin{enumerate}
\item{Post-mergers do not exhibit a statistically significant excess of X-ray AGN over non-merger controls ($p=0.26$), showing an excess of $2.22^{+4.44}_{-2.22}$, where the uncertainties contain the  90\% confidence interval.}
\item{However, for varying AGN mid-IR colour thresholds, these post-mergers exhibit an AGN excess of $\sim17$ or more ($p=0.001$), implying that post-mergers more frequently host intrinsically luminous AGN. The disparity between the X-ray and mid-IR AGN excesses in post-mergers is highly significant ($p=2.7\times10^{-4}$).}
\item{Post-mergers exhibit an optical AGN excess of $\sim3$ ($p\lesssim0.01$), significantly lower than the mid-IR excess, consistent with previous studies and within the 90\% confidence interval of the X-ray excess. By number, however, optical AGN are more common than mid-IR AGN; this holds in our post-merger sample as it does in the general AGN population. Indeed, the majority of post-mergers in our sample host optical AGN, implying that while luminous, bolometrically-dominant AGN are preferentially hosted in late-stage galaxy mergers, lower-level AGN activity is still more prevalent overall.}
\end{enumerate}

\noindent We emphasise that this study is limited by the small numbers of post-mergers available with \textit{XMM-Newton} observations, and we have taken care to calculate statistical significance as robustly as possible and to not over-state our conclusions. Nonetheless, our results are consistent with the picture of mergers driving gas towards the centres of galaxies, fueling both star formation and AGN activity, with the heaviest AGN obscuration and the highest AGN luminosities occurring more frequently in the post-merger stage.

\section*{Acknowledgements}

We thank the anonymous referee for their thorough review that greatly improved this work.

This research made use of Astropy,\footnote{\url{http://www.astropy.org}} a community-developed core Python package for Astronomy \citep{2013A&A...558A..33A, 2018AJ....156..123A}, as well as TOPCAT \citep{2005ASPC..347...29T},\footnote{\url{http://www.star.bris.ac.uk/~mbt/topcat/}} version 4.7.

Funding for the SDSS and SDSS-II has been provided by the Alfred P. Sloan Foundation, the Participating Institutions, the National Science Foundation, the U.S. Department of Energy, the National Aeronautics and Space Administration, the Japanese Monbukagakusho, the Max Planck Society, and the Higher Education Funding Council for England. The SDSS Web Site is http://www.sdss.org/.

The SDSS is managed by the Astrophysical Research Consortium for the Participating Institutions. The Participating Institutions are the American Museum of Natural History, Astrophysical Institute Potsdam, University of Basel, University of Cambridge, Case Western Reserve University, University of Chicago, Drexel University, Fermilab, the Institute for Advanced Study, the Japan Participation Group, Johns Hopkins University, the Joint Institute for Nuclear Astrophysics, the Kavli Institute for Particle Astrophysics and Cosmology, the Korean Scientist Group, the Chinese Academy of Sciences (LAMOST), Los Alamos National Laboratory, the Max-Planck-Institute for Astronomy (MPIA), the Max-Planck-Institute for Astrophysics (MPA), New Mexico State University, Ohio State University, University of Pittsburgh, University of Portsmouth, Princeton University, the United States Naval Observatory, and the University of Washington.

L.B.\ acknowledges support from NSF grant AST-1715413.



\bibliographystyle{mnras}
\bibliography{secrest+20_R1} 

\begin{thebibliography}{}
\makeatletter
\relax
\def\mn@urlcharsother{\let\do\@makeother \do\$\do\&\do\#\do\^\do\_\do\%\do\~}
\def\mn@doi{\begingroup\mn@urlcharsother \@ifnextchar [ {\mn@doi@}
  {\mn@doi@[]}}
\def\mn@doi@[#1]#2{\def\@tempa{#1}\ifx\@tempa\@empty \href
  {http://dx.doi.org/#2} {doi:#2}\else \href {http://dx.doi.org/#2} {#1}\fi
  \endgroup}
\def\mn@eprint#1#2{\mn@eprint@#1:#2::\@nil}
\def\mn@eprint@arXiv#1{\href {http://arxiv.org/abs/#1} {{\tt arXiv:#1}}}
\def\mn@eprint@dblp#1{\href {http://dblp.uni-trier.de/rec/bibtex/#1.xml}
  {dblp:#1}}
\def\mn@eprint@#1:#2:#3:#4\@nil{\def\@tempa {#1}\def\@tempb {#2}\def\@tempc
  {#3}\ifx \@tempc \@empty \let \@tempc \@tempb \let \@tempb \@tempa \fi \ifx
  \@tempb \@empty \def\@tempb {arXiv}\fi \@ifundefined
  {mn@eprint@\@tempb}{\@tempb:\@tempc}{\expandafter \expandafter \csname
  mn@eprint@\@tempb\endcsname \expandafter{\@tempc}}}

\bibitem[\protect\citeauthoryear{{Abazajian} et~al.,}{{Abazajian}
  et~al.}{2009}]{2009ApJS..182..543A}
{Abazajian} K.~N.,  et~al., 2009, \mn@doi [\apjs]
  {10.1088/0067-0049/182/2/543}, \href
  {https://ui.adsabs.harvard.edu/abs/2009ApJS..182..543A} {182, 543}

\bibitem[\protect\citeauthoryear{{Asmus}, {Gandhi}, {H{\"o}nig}, {Smette}  \&
  {Duschl}}{{Asmus} et~al.}{2015}]{2015MNRAS.454..766A}
{Asmus} D.,  {Gandhi} P.,  {H{\"o}nig} S.~F.,  {Smette} A.,   {Duschl} W.~J.,
  2015, \mn@doi [\mnras] {10.1093/mnras/stv1950}, \href
  {https://ui.adsabs.harvard.edu/abs/2015MNRAS.454..766A} {454, 766}

\bibitem[\protect\citeauthoryear{{Assef} et~al.,}{{Assef}
  et~al.}{2010}]{2010ApJ...713..970A}
{Assef} R.~J.,  et~al., 2010, \mn@doi [\apj] {10.1088/0004-637X/713/2/970},
  \href {https://ui.adsabs.harvard.edu/abs/2010ApJ...713..970A} {713, 970}

\bibitem[\protect\citeauthoryear{{Astropy Collaboration} et~al.,}{{Astropy
  Collaboration} et~al.}{2013}]{2013A&A...558A..33A}
{Astropy Collaboration} et~al., 2013, \mn@doi [\aap]
  {10.1051/0004-6361/201322068}, \href
  {https://ui.adsabs.harvard.edu/abs/2013A&A...558A..33A} {558, A33}

\bibitem[\protect\citeauthoryear{{Astropy Collaboration} et~al.,}{{Astropy
  Collaboration} et~al.}{2018}]{2018AJ....156..123A}
{Astropy Collaboration} et~al., 2018, \mn@doi [\aj] {10.3847/1538-3881/aabc4f},
  \href {https://ui.adsabs.harvard.edu/abs/2018AJ....156..123A} {156, 123}

\bibitem[\protect\citeauthoryear{{Ba{\~n}ados} et~al.,}{{Ba{\~n}ados}
  et~al.}{2018}]{2018Natur.553..473B}
{Ba{\~n}ados} E.,  et~al., 2018, \mn@doi [\nat] {10.1038/nature25180}, \href
  {https://ui.adsabs.harvard.edu/abs/2018Natur.553..473B} {553, 473}

\bibitem[\protect\citeauthoryear{{Barthelmy} et~al.,}{{Barthelmy}
  et~al.}{2005}]{2005SSRv..120..143B}
{Barthelmy} S.~D.,  et~al., 2005, \mn@doi [\ssr] {10.1007/s11214-005-5096-3},
  \href {https://ui.adsabs.harvard.edu/abs/2005SSRv..120..143B} {120, 143}

\bibitem[\protect\citeauthoryear{{Becker}, {White}  \& {Helfand}}{{Becker}
  et~al.}{1995}]{1995ApJ...450..559B}
{Becker} R.~H.,  {White} R.~L.,   {Helfand} D.~J.,  1995, \mn@doi [\apj]
  {10.1086/176166}, \href
  {https://ui.adsabs.harvard.edu/abs/1995ApJ...450..559B} {450, 559}

\bibitem[\protect\citeauthoryear{{Blecha}, {Snyder}, {Satyapal}  \&
  {Ellison}}{{Blecha} et~al.}{2018}]{2018MNRAS.478.3056B}
{Blecha} L.,  {Snyder} G.~F.,  {Satyapal} S.,   {Ellison} S.~L.,  2018, \mn@doi
  [\mnras] {10.1093/mnras/sty1274}, \href
  {https://ui.adsabs.harvard.edu/abs/2018MNRAS.478.3056B} {478, 3056}

\bibitem[\protect\citeauthoryear{{Brinchmann}, {Charlot}, {White}, {Tremonti},
  {Kauffmann}, {Heckman}  \& {Brinkmann}}{{Brinchmann}
  et~al.}{2004}]{2004MNRAS.351.1151B}
{Brinchmann} J.,  {Charlot} S.,  {White} S.~D.~M.,  {Tremonti} C.,  {Kauffmann}
  G.,  {Heckman} T.,   {Brinkmann} J.,  2004, \mn@doi [\mnras]
  {10.1111/j.1365-2966.2004.07881.x}, \href
  {https://ui.adsabs.harvard.edu/abs/2004MNRAS.351.1151B} {351, 1151}

\bibitem[\protect\citeauthoryear{{Chen} et~al.,}{{Chen}
  et~al.}{2019}]{2019MNRAS.489..855C}
{Chen} J.,  et~al., 2019, \mn@doi [\mnras] {10.1093/mnras/stz2183}, \href
  {https://ui.adsabs.harvard.edu/abs/2019MNRAS.489..855C} {489, 855}

\bibitem[\protect\citeauthoryear{{Chiaberge}, {Gilli}, {Lotz}  \&
  {Norman}}{{Chiaberge} et~al.}{2015}]{2015ApJ...806..147C}
{Chiaberge} M.,  {Gilli} R.,  {Lotz} J.~M.,   {Norman} C.,  2015, \mn@doi
  [\apj] {10.1088/0004-637X/806/2/147}, \href
  {https://ui.adsabs.harvard.edu/abs/2015ApJ...806..147C} {806, 147}

\bibitem[\protect\citeauthoryear{{Cisternas} et~al.,}{{Cisternas}
  et~al.}{2011}]{2011ApJ...726...57C}
{Cisternas} M.,  et~al., 2011, \mn@doi [\apj] {10.1088/0004-637X/726/2/57},
  \href {https://ui.adsabs.harvard.edu/abs/2011ApJ...726...57C} {726, 57}

\bibitem[\protect\citeauthoryear{{Condon}, {Cotton}, {Greisen}, {Yin},
  {Perley}, {Taylor}  \& {Broderick}}{{Condon}
  et~al.}{1998}]{1998AJ....115.1693C}
{Condon} J.~J.,  {Cotton} W.~D.,  {Greisen} E.~W.,  {Yin} Q.~F.,  {Perley}
  R.~A.,  {Taylor} G.~B.,   {Broderick} J.~J.,  1998, \mn@doi [\aj]
  {10.1086/300337}, \href
  {https://ui.adsabs.harvard.edu/abs/1998AJ....115.1693C} {115, 1693}

\bibitem[\protect\citeauthoryear{{Dietrich} et~al.,}{{Dietrich}
  et~al.}{2018}]{2018MNRAS.480.3562D}
{Dietrich} J.,  et~al., 2018, \mn@doi [\mnras] {10.1093/mnras/sty2056}, \href
  {https://ui.adsabs.harvard.edu/abs/2018MNRAS.480.3562D} {480, 3562}

\bibitem[\protect\citeauthoryear{{Donley} et~al.,}{{Donley}
  et~al.}{2012}]{2012ApJ...748..142D}
{Donley} J.~L.,  et~al., 2012, \mn@doi [\apj] {10.1088/0004-637X/748/2/142},
  \href {https://ui.adsabs.harvard.edu/abs/2012ApJ...748..142D} {748, 142}

\bibitem[\protect\citeauthoryear{{Donley} et~al.,}{{Donley}
  et~al.}{2018}]{2018ApJ...853...63D}
{Donley} J.~L.,  et~al., 2018, \mn@doi [\apj] {10.3847/1538-4357/aa9ffa}, \href
  {https://ui.adsabs.harvard.edu/abs/2018ApJ...853...63D} {853, 63}

\bibitem[\protect\citeauthoryear{{Ellison}, {Patton}, {Mendel}  \&
  {Scudder}}{{Ellison} et~al.}{2011}]{2011MNRAS.418.2043E}
{Ellison} S.~L.,  {Patton} D.~R.,  {Mendel} J.~T.,   {Scudder} J.~M.,  2011,
  \mn@doi [\mnras] {10.1111/j.1365-2966.2011.19624.x}, \href
  {https://ui.adsabs.harvard.edu/abs/2011MNRAS.418.2043E} {418, 2043}

\bibitem[\protect\citeauthoryear{{Ellison}, {Mendel}, {Patton}  \&
  {Scudder}}{{Ellison} et~al.}{2013}]{2013MNRAS.435.3627E}
{Ellison} S.~L.,  {Mendel} J.~T.,  {Patton} D.~R.,   {Scudder} J.~M.,  2013,
  \mn@doi [\mnras] {10.1093/mnras/stt1562}, \href
  {https://ui.adsabs.harvard.edu/abs/2013MNRAS.435.3627E} {435, 3627}

\bibitem[\protect\citeauthoryear{{Ellison}, {Patton}  \& {Hickox}}{{Ellison}
  et~al.}{2015}]{2015MNRAS.451L..35E}
{Ellison} S.~L.,  {Patton} D.~R.,   {Hickox} R.~C.,  2015, \mn@doi [\mnras]
  {10.1093/mnrasl/slv061}, \href
  {https://ui.adsabs.harvard.edu/abs/2015MNRAS.451L..35E} {451, L35}

\bibitem[\protect\citeauthoryear{{Ellison}, {Viswanathan}, {Patton},
  {Bottrell}, {McConnachie}, {Gwyn}  \& {Cuillandre}}{{Ellison}
  et~al.}{2019}]{2019MNRAS.487.2491E}
{Ellison} S.~L.,  {Viswanathan} A.,  {Patton} D.~R.,  {Bottrell} C.,
  {McConnachie} A.~W.,  {Gwyn} S.,   {Cuillandre} J.-C.,  2019, \mn@doi
  [\mnras] {10.1093/mnras/stz1431}, \href
  {https://ui.adsabs.harvard.edu/abs/2019MNRAS.487.2491E} {487, 2491}

\bibitem[\protect\citeauthoryear{{Fan} et~al.,}{{Fan}
  et~al.}{2016}]{2016ApJ...822L..32F}
{Fan} L.,  et~al., 2016, \mn@doi [\apjl] {10.3847/2041-8205/822/2/L32}, \href
  {https://ui.adsabs.harvard.edu/abs/2016ApJ...822L..32F} {822, L32}

\bibitem[\protect\citeauthoryear{{Fornasini}, {Civano}, {Fabbiano}, {Elvis},
  {Marchesi}, {Miyaji}  \& {Zezas}}{{Fornasini}
  et~al.}{2018}]{2018ApJ...865...43F}
{Fornasini} F.~M.,  {Civano} F.,  {Fabbiano} G.,  {Elvis} M.,  {Marchesi} S.,
  {Miyaji} T.,   {Zezas} A.,  2018, \mn@doi [\apj] {10.3847/1538-4357/aada4e},
  \href {https://ui.adsabs.harvard.edu/abs/2018ApJ...865...43F} {865, 43}

\bibitem[\protect\citeauthoryear{{Gehrels}}{{Gehrels}}{1986}]{1986ApJ...303..336G}
{Gehrels} N.,  1986, \mn@doi [\apj] {10.1086/164079}, \href
  {https://ui.adsabs.harvard.edu/abs/1986ApJ...303..336G} {303, 336}

\bibitem[\protect\citeauthoryear{{Glikman}, {Simmons}, {Mailly}, {Schawinski},
  {Urry}  \& {Lacy}}{{Glikman} et~al.}{2015}]{2015ApJ...806..218G}
{Glikman} E.,  {Simmons} B.,  {Mailly} M.,  {Schawinski} K.,  {Urry} C.~M.,
  {Lacy} M.,  2015, \mn@doi [\apj] {10.1088/0004-637X/806/2/218}, \href
  {https://ui.adsabs.harvard.edu/abs/2015ApJ...806..218G} {806, 218}

\bibitem[\protect\citeauthoryear{{Goulding} et~al.,}{{Goulding}
  et~al.}{2018}]{2018PASJ...70S..37G}
{Goulding} A.~D.,  et~al., 2018, \mn@doi [\pasj] {10.1093/pasj/psx135}, \href
  {https://ui.adsabs.harvard.edu/abs/2018PASJ...70S..37G} {70, S37}

\bibitem[\protect\citeauthoryear{{Kauffmann} et~al.,}{{Kauffmann}
  et~al.}{2003a}]{2003MNRAS.341...33K}
{Kauffmann} G.,  et~al., 2003a, \mn@doi [\mnras]
  {10.1046/j.1365-8711.2003.06291.x}, \href
  {https://ui.adsabs.harvard.edu/abs/2003MNRAS.341...33K} {341, 33}

\bibitem[\protect\citeauthoryear{{Kauffmann} et~al.,}{{Kauffmann}
  et~al.}{2003b}]{2003MNRAS.346.1055K}
{Kauffmann} G.,  et~al., 2003b, \mn@doi [\mnras]
  {10.1111/j.1365-2966.2003.07154.x}, \href
  {https://ui.adsabs.harvard.edu/abs/2003MNRAS.346.1055K} {346, 1055}

\bibitem[\protect\citeauthoryear{{Kewley}, {Dopita}, {Sutherland}, {Heisler}
  \& {Trevena}}{{Kewley} et~al.}{2001}]{2001ApJ...556..121K}
{Kewley} L.~J.,  {Dopita} M.~A.,  {Sutherland} R.~S.,  {Heisler} C.~A.,
  {Trevena} J.,  2001, \mn@doi [\apj] {10.1086/321545}, \href
  {https://ui.adsabs.harvard.edu/abs/2001ApJ...556..121K} {556, 121}

\bibitem[\protect\citeauthoryear{{Kocevski} et~al.,}{{Kocevski}
  et~al.}{2012}]{2012ApJ...744..148K}
{Kocevski} D.~D.,  et~al., 2012, \mn@doi [\apj] {10.1088/0004-637X/744/2/148},
  \href {https://ui.adsabs.harvard.edu/abs/2012ApJ...744..148K} {744, 148}

\bibitem[\protect\citeauthoryear{{Kocevski} et~al.,}{{Kocevski}
  et~al.}{2015}]{2015ApJ...814..104K}
{Kocevski} D.~D.,  et~al., 2015, \mn@doi [\apj] {10.1088/0004-637X/814/2/104},
  \href {https://ui.adsabs.harvard.edu/abs/2015ApJ...814..104K} {814, 104}

\bibitem[\protect\citeauthoryear{{Kormendy} \& {Ho}}{{Kormendy} \&
  {Ho}}{2013}]{2013ARA&A..51..511K}
{Kormendy} J.,  {Ho} L.~C.,  2013, \mn@doi [\araa]
  {10.1146/annurev-astro-082708-101811}, \href
  {https://ui.adsabs.harvard.edu/abs/2013ARA&A..51..511K} {51, 511}

\bibitem[\protect\citeauthoryear{{Koss}, {Mushotzky}, {Veilleux}  \&
  {Winter}}{{Koss} et~al.}{2010}]{2010ApJ...716L.125K}
{Koss} M.,  {Mushotzky} R.,  {Veilleux} S.,   {Winter} L.,  2010, \mn@doi
  [\apjl] {10.1088/2041-8205/716/2/L125}, \href
  {https://ui.adsabs.harvard.edu/abs/2010ApJ...716L.125K} {716, L125}

\bibitem[\protect\citeauthoryear{{Koss}, {Mushotzky}, {Veilleux}, {Winter},
  {Baumgartner}, {Tueller}, {Gehrels}  \& {Valencic}}{{Koss}
  et~al.}{2011}]{2011ApJ...739...57K}
{Koss} M.,  {Mushotzky} R.,  {Veilleux} S.,  {Winter} L.~M.,  {Baumgartner} W.,
   {Tueller} J.,  {Gehrels} N.,   {Valencic} L.,  2011, \mn@doi [\apj]
  {10.1088/0004-637X/739/2/57}, \href
  {https://ui.adsabs.harvard.edu/abs/2011ApJ...739...57K} {739, 57}

\bibitem[\protect\citeauthoryear{{Koss}, {Mushotzky}, {Treister}, {Veilleux},
  {Vasudevan}  \& {Trippe}}{{Koss} et~al.}{2012}]{2012ApJ...746L..22K}
{Koss} M.,  {Mushotzky} R.,  {Treister} E.,  {Veilleux} S.,  {Vasudevan} R.,
  {Trippe} M.,  2012, \mn@doi [\apjl] {10.1088/2041-8205/746/2/L22}, \href
  {https://ui.adsabs.harvard.edu/abs/2012ApJ...746L..22K} {746, L22}

\bibitem[\protect\citeauthoryear{{Koss} et~al.,}{{Koss}
  et~al.}{2018}]{2018Natur.563..214K}
{Koss} M.~J.,  et~al., 2018, \mn@doi [\nat] {10.1038/s41586-018-0652-7}, \href
  {https://ui.adsabs.harvard.edu/abs/2018Natur.563..214K} {563, 214}

\bibitem[\protect\citeauthoryear{{Lackner} et~al.,}{{Lackner}
  et~al.}{2014}]{2014AJ....148..137L}
{Lackner} C.~N.,  et~al., 2014, \mn@doi [\aj] {10.1088/0004-6256/148/6/137},
  \href {https://ui.adsabs.harvard.edu/abs/2014AJ....148..137L} {148, 137}

\bibitem[\protect\citeauthoryear{{Lanzuisi} et~al.,}{{Lanzuisi}
  et~al.}{2015}]{2015A&A...573A.137L}
{Lanzuisi} G.,  et~al., 2015, \mn@doi [\aap] {10.1051/0004-6361/201424924},
  \href {https://ui.adsabs.harvard.edu/abs/2015A&A...573A.137L} {573, A137}

\bibitem[\protect\citeauthoryear{{Lehmer}, {Alexander}, {Bauer}, {Brandt},
  {Goulding}, {Jenkins}, {Ptak}  \& {Roberts}}{{Lehmer}
  et~al.}{2010}]{2010ApJ...724..559L}
{Lehmer} B.~D.,  {Alexander} D.~M.,  {Bauer} F.~E.,  {Brandt} W.~N.,
  {Goulding} A.~D.,  {Jenkins} L.~P.,  {Ptak} A.,   {Roberts} T.~P.,  2010,
  \mn@doi [\apj] {10.1088/0004-637X/724/1/559}, \href
  {https://ui.adsabs.harvard.edu/abs/2010ApJ...724..559L} {724, 559}

\bibitem[\protect\citeauthoryear{{Lehmer} et~al.,}{{Lehmer}
  et~al.}{2019}]{2019ApJS..243....3L}
{Lehmer} B.~D.,  et~al., 2019, \mn@doi [\apjs] {10.3847/1538-4365/ab22a8},
  \href {https://ui.adsabs.harvard.edu/abs/2019ApJS..243....3L} {243, 3}

\bibitem[\protect\citeauthoryear{{Lintott} et~al.,}{{Lintott}
  et~al.}{2008}]{2008MNRAS.389.1179L}
{Lintott} C.~J.,  et~al., 2008, \mn@doi [\mnras]
  {10.1111/j.1365-2966.2008.13689.x}, \href
  {https://ui.adsabs.harvard.edu/abs/2008MNRAS.389.1179L} {389, 1179}

\bibitem[\protect\citeauthoryear{{Lintott} et~al.,}{{Lintott}
  et~al.}{2011}]{2011MNRAS.410..166L}
{Lintott} C.,  et~al., 2011, \mn@doi [\mnras]
  {10.1111/j.1365-2966.2010.17432.x}, \href
  {https://ui.adsabs.harvard.edu/abs/2011MNRAS.410..166L} {410, 166}

\bibitem[\protect\citeauthoryear{{Powell} et~al.,}{{Powell}
  et~al.}{2018}]{2018ApJ...858..110P}
{Powell} M.~C.,  et~al., 2018, \mn@doi [\apj] {10.3847/1538-4357/aabd7f}, \href
  {https://ui.adsabs.harvard.edu/abs/2018ApJ...858..110P} {858, 110}

\bibitem[\protect\citeauthoryear{{Ricci} et~al.,}{{Ricci}
  et~al.}{2017}]{2017MNRAS.468.1273R}
{Ricci} C.,  et~al., 2017, \mn@doi [\mnras] {10.1093/mnras/stx173}, \href
  {https://ui.adsabs.harvard.edu/abs/2017MNRAS.468.1273R} {468, 1273}

\bibitem[\protect\citeauthoryear{{Salim} et~al.,}{{Salim}
  et~al.}{2007}]{2007ApJS..173..267S}
{Salim} S.,  et~al., 2007, \mn@doi [\apjs] {10.1086/519218}, \href
  {https://ui.adsabs.harvard.edu/abs/2007ApJS..173..267S} {173, 267}

\bibitem[\protect\citeauthoryear{{Satyapal}, {Ellison}, {McAlpine}, {Hickox},
  {Patton}  \& {Mendel}}{{Satyapal} et~al.}{2014}]{2014MNRAS.441.1297S}
{Satyapal} S.,  {Ellison} S.~L.,  {McAlpine} W.,  {Hickox} R.~C.,  {Patton}
  D.~R.,   {Mendel} J.~T.,  2014, \mn@doi [\mnras] {10.1093/mnras/stu650},
  \href {https://ui.adsabs.harvard.edu/abs/2014MNRAS.441.1297S} {441, 1297}

\bibitem[\protect\citeauthoryear{{Satyapal} et~al.,}{{Satyapal}
  et~al.}{2017}]{2017ApJ...848..126S}
{Satyapal} S.,  et~al., 2017, \mn@doi [\apj] {10.3847/1538-4357/aa88ca}, \href
  {https://ui.adsabs.harvard.edu/abs/2017ApJ...848..126S} {848, 126}

\bibitem[\protect\citeauthoryear{{Satyapal}, {Abel}  \& {Secrest}}{{Satyapal}
  et~al.}{2018}]{2018ApJ...858...38S}
{Satyapal} S.,  {Abel} N.~P.,   {Secrest} N.~J.,  2018, \mn@doi [\apj]
  {10.3847/1538-4357/aab7f8}, \href
  {https://ui.adsabs.harvard.edu/abs/2018ApJ...858...38S} {858, 38}

\bibitem[\protect\citeauthoryear{{Schawinski}, {Simmons}, {Urry}, {Treister}
  \& {Glikman}}{{Schawinski} et~al.}{2012}]{2012MNRAS.425L..61S}
{Schawinski} K.,  {Simmons} B.~D.,  {Urry} C.~M.,  {Treister} E.,   {Glikman}
  E.,  2012, \mn@doi [\mnras] {10.1111/j.1745-3933.2012.01302.x}, \href
  {https://ui.adsabs.harvard.edu/abs/2012MNRAS.425L..61S} {425, L61}

\bibitem[\protect\citeauthoryear{{Secrest} et~al.,}{{Secrest}
  et~al.}{2015}]{2015ApJ...798...38S}
{Secrest} N.~J.,  et~al., 2015, \mn@doi [\apj] {10.1088/0004-637X/798/1/38},
  \href {https://ui.adsabs.harvard.edu/abs/2015ApJ...798...38S} {798, 38}

\bibitem[\protect\citeauthoryear{{Stasi{\'n}ska}, {Cid Fernandes}, {Mateus},
  {Sodr{\'e}}  \& {Asari}}{{Stasi{\'n}ska} et~al.}{2006}]{2006MNRAS.371..972S}
{Stasi{\'n}ska} G.,  {Cid Fernandes} R.,  {Mateus} A.,  {Sodr{\'e}} L.,
  {Asari} N.~V.,  2006, \mn@doi [\mnras] {10.1111/j.1365-2966.2006.10732.x},
  \href {https://ui.adsabs.harvard.edu/abs/2006MNRAS.371..972S} {371, 972}

\bibitem[\protect\citeauthoryear{{Stern}}{{Stern}}{2015}]{2015ApJ...807..129S}
{Stern} D.,  2015, \mn@doi [\apj] {10.1088/0004-637X/807/2/129}, \href
  {https://ui.adsabs.harvard.edu/abs/2015ApJ...807..129S} {807, 129}

\bibitem[\protect\citeauthoryear{{Stern} et~al.,}{{Stern}
  et~al.}{2012}]{2012ApJ...753...30S}
{Stern} D.,  et~al., 2012, \mn@doi [\apj] {10.1088/0004-637X/753/1/30}, \href
  {https://ui.adsabs.harvard.edu/abs/2012ApJ...753...30S} {753, 30}

\bibitem[\protect\citeauthoryear{{Taylor}}{{Taylor}}{2005}]{2005ASPC..347...29T}
{Taylor} M.~B.,  2005, in {Shopbell} P.,  {Britton} M.,   {Ebert} R.,  eds,
  Astronomical Society of the Pacific Conference Series Vol. 347, Astronomical
  Data Analysis Software and Systems XIV. p.~29

\bibitem[\protect\citeauthoryear{{Treister}, {Schawinski}, {Urry}  \&
  {Simmons}}{{Treister} et~al.}{2012}]{2012ApJ...758L..39T}
{Treister} E.,  {Schawinski} K.,  {Urry} C.~M.,   {Simmons} B.~D.,  2012,
  \mn@doi [\apjl] {10.1088/2041-8205/758/2/L39}, \href
  {https://ui.adsabs.harvard.edu/abs/2012ApJ...758L..39T} {758, L39}

\bibitem[\protect\citeauthoryear{{Trump} et~al.,}{{Trump}
  et~al.}{2015}]{2015ApJ...811...26T}
{Trump} J.~R.,  et~al., 2015, \mn@doi [\apj] {10.1088/0004-637X/811/1/26},
  \href {https://ui.adsabs.harvard.edu/abs/2015ApJ...811...26T} {811, 26}

\bibitem[\protect\citeauthoryear{{Villforth} et~al.,}{{Villforth}
  et~al.}{2014}]{2014MNRAS.439.3342V}
{Villforth} C.,  et~al., 2014, \mn@doi [\mnras] {10.1093/mnras/stu173}, \href
  {https://ui.adsabs.harvard.edu/abs/2014MNRAS.439.3342V} {439, 3342}

\bibitem[\protect\citeauthoryear{{Villforth} et~al.,}{{Villforth}
  et~al.}{2017}]{2017MNRAS.466..812V}
{Villforth} C.,  et~al., 2017, \mn@doi [\mnras] {10.1093/mnras/stw3037}, \href
  {https://ui.adsabs.harvard.edu/abs/2017MNRAS.466..812V} {466, 812}

\bibitem[\protect\citeauthoryear{{Villforth}, {Herbst}, {Hamann}, {Hamilton},
  {Bertemes}, {Efthymiadou}  \& {Hewlett}}{{Villforth}
  et~al.}{2019}]{2019MNRAS.483.2441V}
{Villforth} C.,  {Herbst} H.,  {Hamann} F.,  {Hamilton} T.,  {Bertemes} C.,
  {Efthymiadou} A.,   {Hewlett} T.,  2019, \mn@doi [\mnras]
  {10.1093/mnras/sty3271}, \href
  {https://ui.adsabs.harvard.edu/abs/2019MNRAS.483.2441V} {483, 2441}

\bibitem[\protect\citeauthoryear{{Voges} et~al.,}{{Voges}
  et~al.}{1999}]{1999A&A...349..389V}
{Voges} W.,  et~al., 1999, \aap, \href
  {https://ui.adsabs.harvard.edu/abs/1999A&A...349..389V} {349, 389}

\bibitem[\protect\citeauthoryear{{Webb} et~al.,}{{Webb}
  et~al.}{2019}]{2019A&A...submitted}
{Webb} N.~A.,  et~al., 2019, \aap, submitted

\bibitem[\protect\citeauthoryear{{Wright} et~al.,}{{Wright}
  et~al.}{2010}]{2010AJ....140.1868W}
{Wright} E.~L.,  et~al., 2010, \mn@doi [\aj] {10.1088/0004-6256/140/6/1868},
  \href {https://ui.adsabs.harvard.edu/abs/2010AJ....140.1868W} {140, 1868}

\makeatother
\end{thebibliography}







\bsp    
\label{lastpage}
\end{document}